\documentclass[10pt,twosided]{article}
\date{}
\usepackage{extsizes}
\usepackage{amsmath}
\usepackage{amsfonts}
\usepackage{amssymb}
\usepackage{graphicx}
\usepackage{braket}
\date{\today}
\begin{document}

\title{{\bf{Exact solution of damped harmonic oscillator with a magnetic field in a time dependent noncommutative space }}}

\author{
{\bf {\normalsize Manjari Dutta}$^{a}
$\thanks{manjaridutta@boson.bose.res.in}},
{\bf {\normalsize Shreemoyee Ganguly}
$^{b}$\thanks{ganguly.shreemoyee@gmail.com}},
{\bf {\normalsize Sunandan Gangopadhyay}
$^{c}$\thanks{ sunandan.gangopadhyay@bose.res.in, sunandan.gangopadhyay@gmail.com}}
\\
$^{a,c}$ {\normalsize Department of Theoretical Sciences},\\
{\normalsize S.N. Bose National Centre for Basic Sciences},\\
{\normalsize JD Block, Sector III, Salt Lake, Kolkata 700106, India}\\
$^{b}$ {\normalsize Department of Basic Science and Humanities,}\\
{\normalsize University of Engineering and Management (UEM),}\\
{\normalsize B/5, Plot No.III, Action Area-III, Newtown, Kolkata 700156}
}
\date{}

\maketitle
\begin{abstract}
\noindent In this paper we have obtained the exact eigenstates of a two dimensional damped harmonic oscillator in the presence of an external magnetic field varying with respect to time in time dependent noncommutative space. It has been observed that for some specific choices of the damping factor, the time dependent frequency of the oscillator and the time dependent external magnetic field, there exists interesting solutions of the time dependent noncommutative parameters following from the solutions of the Ermakov-Pinney equation. Further, these solutions enable us to get exact analytic forms for the phase which relates the eigenstates of the Hamiltonian with the eigenstates of the Lewis invariant. Then we compute the expectation value of the Hamiltonian. The expectation values of the energy are found to vary with time for different solutions of the Ermakov-Pinney equation corresponding to different choices of the damping factor, the time dependent frequency of the oscillator and the time dependent applied magnetic field. We also compare our results with those in the
absence of the magnetic field obtained earlier.

\end{abstract}

\vskip 1cm
\newpage
\section{Introduction}
The Landau problem of a charged particle moving in a two dimensional plane 
under the influence of a magnetic field acting perpendicular to the plane has 
been looked upon by physicists over the years, not only from the pedagogical 
interest of formation of discrete energy levels known as Landau levels, but 
also due to the multifaceted applications of this problem. In this context, 
an extremely intriguing 
problem is that of a charged oscillator placed in a magnetic field, 
acting perpendicular to the plane in which it is oscillating.  
In addition, one may also consider an electric field lying along the plane of oscillation. The eigenfunction and the eigenvalues of a charged particle in a 
magnetic field in the presence of a 
time-dependent background electric field with time-dependent mass and frequency have been determined in \cite{lawson}. The
problem becomes even more fascinating when one places the Landau oscillator in a noncommutative phase space. 
This problem 
has been looked upon by an earlier study~\cite{gouba} in the presence of a 
time dependent magnetic field. The system there was considered in a noncommutative 
space. The simplest setting of noncommutative [NC] space is a two dimensional quantum mechanical space in which one replaces the standard set of commutation relations between the
canonical coordinates by NC commutation relations $[X, Y]=i\theta$, where $\theta$ is a positive real constant. Study of quantum mechanical systems in such NC 
space has allured theoretical physicists since the work by Synder~\cite{Synder}. 
Since then the necessity for NC spaces has been established to ensure the attainment of 
gravitational stability~\cite{Doplicher} in the present theories of quantum 
gravity, namely, string theory \cite{amati, sw} and loop quantum gravity \cite{rov}. This has triggered several studies on quantum mechanical systems in such spaces in the literature \cite{suss}-\cite{fgs}.

However, the mentioned study~\cite{gouba} on charged oscillators in the 
presence of magnetic field, considers only noncommutativity amongst position 
variables. So, we in our present communication extend the study to a 
space where noncommutativity exists not only amongst spatial variables but 
also amongst momentum variables. Also unlike the previous study we have considered the NC space to be time dependent. Moreover, our oscillator is considered 
to be damped by an explicit damping factor in order to model a realistic 
situation. A damped oscillator in two dimensional NC space has been studied by us in an 
earlier communication~\cite{SG}. Before our communication one of the very few 
works which had studied damped quantum harmonic oscillators in two dimensional 
space was the work by Lawson 
{\it et.al.}~\cite{Lawson}. We extended their model to a two dimensional NC 
space. But at present our objective is to study how 
the interplay of damping and an external time dependent magnetic field 
modulates the energetics of a charged oscillator in a time dependent NC space 
where spatial as well as momentum noncommutativity is present.

Our present study is one of the very first to study a damped quantum harmonic oscillator with time varying frequency in two dimensional NC space in the presence of a time-dependent magnetic field. In our earlier study~\cite{SG} we had seen that the expectation value 
of energy of an oscillator decays due to damping even in NC space. In the present study we 
intend to investigate the change in energetics of the damped quantum 
oscillator in NC space under the influence of magnetic field having various 
kinds of time dependence. For this purpose we first set up the Hamiltonian 
of the damped quantum harmonic oscillator in two dimensional NC space under the 
presence of a time varying magnetic field and then express it in terms of 
commutative variables. This is done in Section 2. After that we solve the Hamiltonian using the method of invariants~\cite{Lewis, Lewis2, Lewis3} in Section 3. It must be 
noted that the eigenfunction of the said Hamiltonian is a product of the 
eigenfunction of the invariant and a phase factor.  Both the eigenfunction and phase factor are expressed in terms 
of time dependent parameters which obey the non-linear differential equation known as Ermakov-Pinney (EP) equation~\cite{Ermakov,Pinney}. Next in Section 4 
we choose the parameters of the damped system in the presence of magnetic field in such a way that they satisfy all the equations representing the system as well as provide us exact closed form solutions for the system for various choices 
of time variation of the applied field. In Section 5 we calculate the 
expectation value of energy of the damped oscillator explicitly and 
graphically explore how the time dependence of the magnetic field alters the 
time evolution of the energetics of the damped quantum system having a time 
varying frequency of oscillation in two dimensional NC space.
  


\section{Model of the two-dimensional harmonic oscillator in magnetic field}
In our study the system that we consider is a combination of two non-interacting damped harmonic oscillators affected by a time dependent magnetic field in two dimensional NC space. Both the oscillators have equal time dependent 
frequencies, coefficients of friction, equal mass and equal charge in 
NC space. Such a model of two-dimensional Landau problem of 
harmonic oscillator in a time dependent magnetic field was 
considered in an earlier communication~\cite{gouba} in spatially noncommutative configuration space. In this work, however, we extend the model by considering the system in time dependent NC space. Also it must be noted that the 
noncommutativity we consider is not restricted to spatial variables like the earlier studies but also extends to momentum noncommutativity.\footnote{We shall be considering NC phase space in our work. However, we shall be generically referring to this as NC space.}

\noindent The Hamiltonian of the two dimensional oscillator in magnetic field 
has the following form,
\begin{equation}
H(t)=\dfrac{f(t)}{2M}\left[(P_1-q A_1)^2+(P_2-q A_2)^2\right]+\dfrac{M\omega^2(t)}{2f(t)}({X_1}^2+{X_2}^2)~~;\label{1}
\end{equation}
where the damping factor $f(t)$ is given by,
\begin{equation}
f(t)=e^{-\int_{0}^t\eta(s)ds} 
\label{1x}
\end{equation}
with $\eta(s)$ being the coefficient of friction and $A_i$,~the vector potential of a time dependent magnetic field $B(t)$ is chosen in Coulomb gauge as,
\begin{equation}
A_i=-\dfrac{B(t)}{2}\epsilon_{ij}X^{j}~~;\label{A}
\end{equation} 
where $i,j=1,2$ and $\epsilon_{ij}=-\epsilon_{ji}$ with $\epsilon_{12}=1$. Here $\omega(t)$ is the 
time dependent angular frequency of the oscillators , $M$ and $q$ are their mass and charge respectively. The position and momentum 
coordinates $(X_i,P_i)$ are noncommuting variables in NC space, that is, 
their commutators are $[X_1,X_2]~\neq~0$ and $[P_1,P_2]~\neq~0$. The 
corresponding canonical variables $(x_i,p_i)$ in commutative space are such 
that the commutator $[x_i,p_j]=i\hbar\delta_{i,j}$, $[x_i,x_j]=0=[p_i,p_j]$; ($i,j=1,2$).

In order to express the NC Hamiltonian in terms of the standard commutative variables explicitly, we apply the standard Bopp-shift relations \cite{mez} ($\hbar=1$): 
\begin{eqnarray}
& X_1=x_1-\dfrac{\theta(t)}{2}p_2\,\,\,;\,\,\,X_2=x_2+\dfrac{\theta(t)}{2}p_1\\
& P_1=p_1+\dfrac{\Omega(t)}{2}x_2\,\,\,;\,\,\,P_2=p_2-\dfrac{\Omega(t)}{2}x_1 \,\,.
\label{eqn1}
\end{eqnarray}
Here $\theta(t)$ and $\Omega(t)$ are the NC parameters for space 
and momentum respectively, such that $[X_1,X_2]~=i\theta(t)$, 
$[P_1,P_2]~=i\Omega(t)$ and $[X_1,P_1]=i[1+\frac{\theta(t)\Omega(t)}{4}]=[X_2,P_2]$; ($X_1\equiv X$, $X_2 \equiv Y$, $P_1\equiv P_x$, $P_2 \equiv P_y$). 

\noindent The Hamiltonian in terms of $(x_i,p_i)$ coordinates is therefore given by the following 
relation,
\begin{equation}
H=\dfrac{a(t)}{2}({p_1}^2+{p_2}^2)+\dfrac{b(t)}{2}({x_1}^2+{x_2}^2)+c(t)({p_1}{x_2}-{p_2}{x_1})\,\,\,.\label{ham}
\end{equation}
The time dependent coefficients in the above Hamiltonian are given as,
\begin{align}
&a(t)=\dfrac{f(t)}{M}+\dfrac{qB(t)f(t)\theta(t)}{2M}+\dfrac{1}{4}\left[\dfrac{q^2B^2(t)f(t)}{4M}+\dfrac{M\omega^2(t)}{f(t)}  \right]\theta^2(t)\label{a}\\
&b(t)=\dfrac{q^2B^2(t)f(t)}{4M}+\dfrac{M\omega^2(t)}{f(t)}+\dfrac{qB(t)f(t)\Omega(t)}{2M}+\dfrac{f(t)\Omega^2(t)}{4M}\label{b}\\
&c(t)=\dfrac{1}{2}\left[\dfrac{qB(t)f(t)}{M} \left( 1+\dfrac{\theta(t)\Omega(t)}{4}\right) +\dfrac{\Omega(t)f(t)}{M}+  \left(\dfrac{q^2B^2(t)f(t)}{4M}+\dfrac{M\omega^2(t)}{f(t)}  \right)\theta(t) \right]~.\label{c}
\end{align}

Here it must be noted that although our Hamiltonian given by Eqn.(\ref{ham}) has the same form as that in 
\cite{Dey} and \cite{SG} to study a system of a two dimensional harmonic oscillator and damped harmonic oscillator in NC space, the time dependent 
Hamiltonian coefficients (given by Eqn(s).(\ref{a}),(\ref{b}),(\ref{c})) have 
a modified form. This is because our present system of damped 
harmonic oscillator is studied in the presence of a time dependent magnetic field in two-dimensional NC space. Thus, both the damping factor $f(t)$  and the magnetic field $B(t)$ modulate and alter
the Hamiltonian coefficients from the form considered in earlier study \cite{Dey} and \cite{SG} respectively. It is relevant to mention that those coefficients also differ from those obtained in \cite{gouba} as the considered noncommutativity in that study is time independent and exists only in the configuration 
space.


\section{Solution of the model Hamiltonian}
In order to find the solutions of the model Hamiltonian $H(t)$ (Eqn.(\ref{ham}))
representing the two-dimensional damped harmonic oscillator with magnetic 
field in 
NC space, we follow the route suggested by Lewis {\it et.al.}~\cite{Lewis} in their work. First we 
construct the time-dependent Hermitian invariant operator $I(t)$ corresponding to our Hamiltonian operator $H(t)$ 
(given by Eqn.(\ref{ham})). This is because if one can solve for the eigenfunctions of $I(t)$, $\phi(x_1,x_2)$, such that,
\begin{equation}
I(t)\phi(x_1,x_2)=\epsilon \phi(x_1,x_2)
\label{eqnegn}
\end{equation}
where $\epsilon$ is an eigenvalue of $I(t)$ corresponding to eigenstate $\phi(x_1,x_2)$, one can obtain the 
eigenstates of $H(t)$, $\psi(x_1,x_2,t)$, using the relation given by Lewis {\it et. al.}~\cite{Lewis} which is as 
follows, 
\begin{equation}
\psi(x_1,x_2,t)=e^{i\Theta(t)}\phi(x_1,x_2)
\label{eqnpsi}
\end{equation}
where the real function $\Theta(t)$ which acts as the phase factor will be discussed in details later. 



\subsection{The Time Dependent Invariant}
Next, following the approach taken by Lewis {\it et.al.}~\cite{Lewis}, we need to construct the operator $I(t)$ which 
is an invariant with respect to time, corresponding to the Hamiltonian $H(t)$, as mentioned earlier, such 
that $I(t)$ satisfies the condition,
\begin{equation}
\dfrac{dI}{dt}=\partial_t{I}+\dfrac{1}{i}[I,H]=0.
\label{eqn4}
\end{equation}
The procedure is to choose the Hermitian invariant $I(t)$ to be of the same homogeneous quadratic form defined by Lewis 
{\it et. al.}~\cite{Lewis} for time-dependent harmonic oscillators. However, since we are dealing with a 
two-dimensional system in the present study, $I(t)$ takes on the following form,
\begin{equation}
I(t)=\alpha(t)({p_1}^2+{p_2}^2)+\beta(t)({x_1}^2+{x_2}^2)+\gamma(t)(x_1{p_1}+p_2{x_2}).
\label{eqn5}
\end{equation}
Here we will consider $\hbar=1$ since we choose to work in natural units. Now, using the form of $I(t)$ defined by 
Eqn.(\ref{eqn5}) in  Eqn.(\ref{eqn4}) and equating the coefficients of the canonical variables, we get the 
following relations,
\begin{eqnarray}
\dot{\alpha}(t)&=&-a(t)\gamma(t)\label{eqn6}\\
\dot{\beta}(t)&=&b(t)\gamma(t)\label{eqn7}\\
\dot{\gamma}(t)&=&2\left[\,b(t)\alpha(t)-\beta(t)a(t)\,\right]
\label{eqn8}
\end{eqnarray}
where dot denotes derivative with respect to time $t$.

\noindent To express the above three time dependent parameters $\alpha$,$\beta$ and $\gamma$ in terms of a single time 
dependent parameter, we parametrize $\alpha(t)=\rho^{2}(t)$. Substituting this in Eqn(s).(\ref{eqn6}, \ref{eqn8}), we 
get the other two parameters in terms of $\rho(t)$ as, 
\begin{eqnarray}
\gamma(t)=-\dfrac{2\rho\dot{\rho}}{a(t)}~~,~~
\beta(t)=\dfrac{1}{a(t)}\left[\dfrac{{\dot{\rho}^2}}{a(t)}+{{\rho}^2}b+\dfrac{\rho\ddot{\rho}}{a(t)}-\dfrac{\rho\dot{\rho}\dot{a}}{a^2} \right].\label{eqn10}
\end{eqnarray}
Now, substituting the value of $\beta$ in Eqn.(\ref{eqn7}), we get a non-linear equation in 
$\rho(t)$ which has the form of the non-linear Ermakov-Pinney (EP) equation with a dissipative 
term~\cite{Dey, Ermakov, Pinney}. The form of the non-linear equation is as follows, 
\begin{equation}
\ddot{\rho}-\dfrac{\dot{a}}{a}\dot{\rho}+ab\rho={\xi^2}\dfrac{a^2}{\rho^3}
\label{eqn11}
\end{equation} 
where ${\xi^2}$ is a constant of integration. This equation has similar form to the EP equation obtained in \cite{Dey}, which is expected since 
our $H(t)$ has the same form as theirs. However, once again it should be mentioned that the explicit form of the time-dependent 
coefficients are different due to the presence of the external magnetic field 
as well as the fact that the oscillator is damped.

\noindent Now, using the EP equation we get a simpler form of $\beta$ as,
\begin{eqnarray}
\beta(t)&=&\dfrac{1}{a(t)}\left[\dfrac{{\dot{\rho}^2}}{a(t)}+\dfrac{{\xi^2}{a(t)}}{\rho^2} \right].
\label{eqnew}
\end{eqnarray}  
\noindent Next, substituting the expressions of $\alpha$, $\beta$ and $\gamma$ in 
Eqn.(\ref{eqn5}), we get the following 
expression for $I(t)$,
\begin{equation}
I(t)=\rho^2({p_1}^2+{p_2}^2)+\left(\dfrac{\dot{\rho}^2}{a^2}+\dfrac{{\xi^2}}{\rho^2}\right)({x_1}^2+{x_2}^2)-\dfrac{2\rho\dot{\rho}}{a}(x_1{p_1}+p_2{x_2}).
\label{eqn12}
\end{equation}
This form of the Lewis invariant in Cartesian coordinate is converted to 
polar coordinate using the same procedure as followed in our previous 
communication~\cite{SG}. The invariant in polar coordinate takes the following 
form, 
\begin{eqnarray}
I(t)=\dfrac{\xi^2}{\rho^2}r^2+\left(\rho{p_r}-\dfrac{\dot{\rho}}{a}r\right)^2+\left({\dfrac{\rho{p_\theta}}{r}}\right)^2-\left({\dfrac{\rho\hbar}{2r}}\right)^2
\label{eqn26}
\end{eqnarray}
where the canonical coordinates in polar representation takes the following 
form,
\begin{eqnarray}
p_r=-i\left({\partial}_r+\dfrac{1}{2r} \right)~~,~~
p_{\theta}=-i{\partial_{\theta}}.
\label{eqn22}
\end{eqnarray}
Now we note from Eqn.(\ref{eqn26}) that the invariant $I(t)$ 
has the same form as that used in \cite{Dey} to study the undamped harmonic oscillator in NC space. The time-dependent coefficients involved in the 
present study however differ due to the presence of external magnetic field 
and damping in our system. Thus, 
we can just borrow the expression of eigenfunction and the phase factors 
from \cite{Dey} for our present system.

\subsection{Eigenfunction and phase factor}
We depict the set of eigenstates of the invariant operator $I(t)$ as $\ket{n,l} $, following the convention in 
\cite{Dey}. Here, $n$ and $l$ are integers such that $n+l\geqslant0$. So we have the condition $l\geqslant-n$. 
Thus, if $l=-n+m$, then $m$ is a positive integer; and the corresponding eigenfunction in polar coordinate system has the following form (restoring $\hbar$), 
\begin{eqnarray}
\phi_{n,m-n}(r,\theta)&=&\braket{r,\theta|n,m-n}\\
&=&\lambda_{n}\dfrac{{(i\sqrt{\hbar}\rho)}^m}{\sqrt{m!}}r^{n-m}e^{i(m-n)\theta-\dfrac{a(t)-i\rho\dot{\rho}}{2a(t)
\hbar{\rho}^2}r^2}U\left(-m,1-m+n,\dfrac{r^2}{\hbar\rho^2} \right)
\label{eqn28}
\end{eqnarray}
where $\lambda_n$ is given by
\begin{eqnarray}
\lambda_n^2=\dfrac{1}{\pi{n!}{(\hbar\rho^2)}^{1+n}}~. 
\label{eqn28lam}
\end{eqnarray}
Here, $U\left(-m,1-m+n,\dfrac{r^2}{\hbar\rho^2} \right)$ is 
Tricomi's confluent hypergeometric function \cite{Arfken, uva} and the eigenfunction $\phi_{n,m-n}(r,\theta)$ satisfies the following 
orthonormality relation,
\begin{equation}
\int_0^{2\pi}d\theta\int_0^{\infty}rdr\phi^{*}_{n,m-n}(r,\theta)\phi_{n^{'},m^{'}-n^{'}}(r,\theta)=\delta_{nn^{'}}\delta_{mm^{'}}.
\label{eqn29}
\end{equation}
Again following \cite{Dey}, the expression of the phase factor $\Theta(t)$ is given by,  
\begin{equation}
\Theta_{\,n\,,\,l}(t)\,=\,(\,n\,+\,l\,)\,\int_0^t \left(c(T)-\dfrac{a(T)}{\rho^2(T)} \right)dT~.
\label{eqn30}
\end{equation}
For a given value of $l=-n+m$, it would be given by \cite{Dey},
\begin{equation}
\Theta_{\,n\,,\,m\,-\,n\,}(t)=m\int_0^t \left(c(T)-\dfrac{a(T)}{\rho^2(T)} \right)dT~.
\label{eqn31}
\end{equation}
We shall use this expression to compute the phase explicitly as a function of time for various physical cases in the subsequent discussion.

\noindent The eigenfunction of the Hamiltonian therefore reads (using Eqn(s).(\ref{eqnpsi}, \ref{eqn28}, \ref{eqn31}))
\begin{eqnarray}
\psi_{n,m-n}(r,\theta,t)&=&e^{i\Theta_{n, m-n}(t)}\phi_{n, m-n}(r,\theta)\nonumber\\
&=&\lambda_{n}\dfrac{{(i\sqrt{\hbar}\rho)}^m}{\sqrt{m!}}\exp{\left[im\int_0^t \left(c(T)-\dfrac{a(T)}{\rho^2(T)} \right)dT \right]}
\nonumber\\
&&\times~r^{n-m}e^{i(m-n)\theta-\dfrac{a(t)-i\rho\dot{\rho}}{2a(t)\hbar{\rho}^2}r^2}U\left(-m,1-m+n,\dfrac{r^2}{\hbar\rho^2} \right).
\label{eqn32}
\end{eqnarray}



\section{Solutions for the noncommutative damped oscillator \\in magnetic field}
In this communication 
we are primarily concerned about the evolution of the solution due to the inclusion of a time dependent magnetic field in the system. 
For this purpose we  want to find the eigenfunctions of 
the corresponding Hamiltonian due to interplay of damping and magnetic field. The various 
kinds of damping in the presence of the applied magnetic field are represented by various forms of the time dependent 
coefficients of the Hamiltonian, namely, $a(t)$, $b(t)$ and $c(t)$. 
However, the various forms must be constructed in such a way that they satisfy 
the non-linear EP equation given by Eqn.(\ref{eqn11}). 
The procedure of this construction of exact analytical solutions is based on the Chiellini integrability condition \cite{chill} and this formalism was followed in \cite{Dey}. 
So, in this communication for various forms of $a(t)$ and $b(t)$, we get the corresponding form of $\rho(t)$ 
using the EP equation together with the Chiellini integrability condition. In the subsequent discussion we shall proceed to obtain solutions of the EP equation for the damped oscillator in a magnetic field considered in NC space.


\subsection{Solution Set-I for Ermakov-Pinney equation : Exponentially \\ decaying solutions } 
\subsubsection{The Solution Set}
The simplest kind of solution set of the EP equation under damping is the 
exponentially decaying set used in \cite{Dey}.~The solution set is given by the following
relations, 
\begin{eqnarray}
a(t)=\sigma e^{-\vartheta{t}}\,\,\,,\,\,\,b(t)=\Delta e^{\vartheta{t}}\,\,\,,\,\,\rho(t)={\mu}e^{-\vartheta{t/2}}\,\,\,\,\,
\label{EPsoln1}
\end{eqnarray}
where $\sigma,\Delta$ and $\mu$ are constants.~Here, $\vartheta$ is any 
positive real number.~Substituting the expressions for $a(t),b(t) \,$and$\, \rho(t)$ in the EP equation, we can easily verify the relation between these constants to be as follows, 
\begin{equation}
\mu^4=\dfrac{4\xi^2{\sigma^2}}{4\sigma\Delta-\vartheta^2}~.
\label{EPreln1}
\end{equation}



\subsubsection{Study of the corresponding eigenfunctions}
We now write down the eigenfunctions of the Hamiltonian for the chosen set of 
time-dependent coefficients. For this purpose we need to choose explicit 
forms of the damping factor $f(t)$ , angular frequency of the oscillator 
$\omega(t)$ and the applied magnetic field $B(t)$. The eigenfunction of the invariant $I(t)$ (which is given by 
Eqn.(\ref{eqn28})) takes on the following form for the solution Set-I:
\begin{eqnarray}
\phi_{n,m-n}(r,\theta)=\lambda_{n}\dfrac{{(i{\mu}e^{-\vartheta{t/2}})}^m}{\sqrt{m!}}    r^{n-m}e^{i(m-n)\theta-\dfrac{2\sigma+i\mu^2\vartheta}{4\sigma\mu^2{e^{-\vartheta{t}}}}r^2}U\left(-m,1-m+n,\dfrac{r^2{e^{\vartheta{t}}}}{\mu^2} \right)
\label{eqn33}
\end{eqnarray}
where $\lambda_n$ is given by
\begin{eqnarray}
\lambda_n^{\,2}\,=\,\dfrac{1}{\pi\,n!\,[\mu^2\,exp\,(-\vartheta{t})]^{1+n}}~.
\label{eqn33lam}
\end{eqnarray}
Next, we proceed to obtain explicit expressions of the phase factors for 
various forms of the damping factor $f(t)$ and the angular frequency $\omega(t)$ of the oscillator.~The value of the applied magnetic field $B(t)$ is also tuned accordingly.\\
At first, we consider the most general form of damping factors and the applied magnetic field which are given as follows,
\begin{eqnarray}
f(t)= e^{-\Gamma\,t}~;~\omega(t)=\omega_0 e^{-\delta t/2}~;~B(t)=B_0\,e^{\Lambda\,t}~;\label{expgen}
\end{eqnarray}
where $\Gamma$ and $\delta$ are non-negative real constants and $\Lambda$ is an arbitrary real constant.~Substituting these relations in Eqn(s).(\ref{a}, \ref{b}), we get the most general form of the time dependent NC parameters as,
\begin{align}
\theta(t)=\dfrac{8Me^{-\Gamma\,t}}{q^2B_0^2e^{2(\Lambda-\Gamma)\,t}+4M^2\omega^2_0\,e^{-\delta\,t}}
 &\left[\sqrt{\dfrac{q^2B_0^2\sigma e^{(2\Lambda-\Gamma-\vartheta)\,t}}{4M}+\omega_0^2\,e^{-\delta\,t}\left(M\sigma\,e^{(\Gamma-\vartheta)t}-1\right)}-\dfrac{qB_0e^{(\Lambda-\Gamma)\,t}}{2M} \right]     \label{expgenNC1}\\
\Omega(t)&=  -qB_0\,e^{\Lambda\,t}+2e^{\Gamma\,t}\sqrt{M\Delta\,e^{(\vartheta-\Gamma)t}-M^2\omega_0^2\,e^{-\delta\,t}}~.                                                                                                           \label{expgenNC2}
\end{align}
In order to get the exact analytical form of the phase factor we choose some suitable special forms of the constants $\vartheta$, $\Gamma$, $\delta$ and 
$\Lambda$.

\noindent {\bf $\langle a \rangle $ Set-I~,~Case I}

Here we set the constants 
\begin{eqnarray}
\vartheta=\Gamma~,~\delta=0~,~\Lambda=0~.
\end{eqnarray}
 So, the parameters can be depicted by the following relations, 
\begin{eqnarray}
f(t)= e^{\,-\Gamma\,t}~;~\omega(t)={\omega_0}~;~B(t)=B_0~.\label{case1}
\end{eqnarray}
Therefore, substituting Eqn.(\ref{case1}) in Eqn(s).(\ref{expgenNC1},\ref{expgenNC2}), the reduced form of the NC parameters for this case are as follows,
\begin{align}
\theta(t)=\dfrac{8Me^{-\Gamma\,t}}{q^2B_0^2e^{-2\Gamma\,t}+4M^2\omega^2_0\,}
 &\left[\sqrt{\dfrac{q^2B_0^2\sigma e^{-2\Gamma\,t}}{4M}+\omega_0^2\,\left(M\sigma\,-1\right)}-\dfrac{qB_0e^{-\Gamma\,t}}{2M} \right]\nonumber\\                      
\Omega(t)&=-qB_0\,+2e^{\Gamma\,t}\sqrt{M\Delta\,-M^2\omega_0^2\,}~~~.                                                                                                           
\end{align}
It can be checked that in the limit $B\rightarrow0$, the expressions for $\theta(t)$ and $\Omega(t)$ reduce 
to those in \cite{SG}.~Substituting these relations in the expression for $c(t)$ in Eqn.(\ref{c}), we 
get,
\begin{eqnarray}
c(t)&=&\dfrac{1}{q^2B_0^2e^{-2\Gamma\,t}+4M^2\omega_0^2} \left[\left(4M^2\omega_0^2+2qB_0e^{-\Gamma\,t}\sqrt{M\Delta-M^2\omega_0^2} \right)\sqrt{\dfrac{q^2B_0^2\sigma e^{-2\Gamma\,t} }{4M}+\omega_0^2(M\sigma-1)}         \right.\nonumber \\
&&\left.          -2qB_0M\omega_0^2e^{-\Gamma\,t}-\dfrac{q^2B_0^2}{M}e^{-2\Gamma\,t}\sqrt{M\Delta-M^2\omega_0^2}              \right]+\sqrt{\dfrac{\Delta}{M}-\omega_0^2}                                                                                ~.
\end{eqnarray}
Substituting the expressions of $a(t)\,,\rho(t)$\, and $c(t)$\, in Eqn.(\ref{eqn31})\,,\,we can get an expression for the phase in a closed form as,
\begin{align}
\small
\Theta_{\,n\,,\,l}(t)\,&=(n+l)\int_0^t \left[c(t)-\dfrac{a}{\rho^2}\right]~dT=(n+l)\left[\sqrt{\dfrac{\Delta}{M}-\omega_0^2}-\dfrac{\sigma}{\mu^2}\right]t\nonumber\\
&+\dfrac{(n+l)\omega_0\sqrt{(M\sigma-1)}}{\Gamma}
\,log\,\dfrac{\omega_0\sqrt{M\sigma-1}\,e^{\Gamma\,t}+\sqrt{\dfrac{q^2B_0^2\sigma}{4M}+\omega_0^2(M\sigma-1)e^{2\Gamma\,t}} }{\omega_0\sqrt{M\sigma-1}+\sqrt{\dfrac{q^2B_0^2\sigma}{4M}+\omega_0^2(M\sigma-1)}}
\nonumber\\
&+\dfrac{(n+l)\omega_0}{\Gamma}
\left[ tan^{-1}\dfrac{\omega_0 e^{\Gamma\,t}}{\sqrt{\dfrac{q^2B_0^2\sigma}{4M}+\omega_0^2(M\sigma-1)e^{2\Gamma\,t}}}-
tan^{-1}\dfrac{\omega_0 }{\sqrt{\dfrac{q^2B_0^2\sigma}{4M}+\omega_0^2(M\sigma-1)}}
\right]\nonumber\\
&+\dfrac{(n+l)\sqrt{M\Delta-M^2\omega_0^2}}{M\Gamma} \left[tanh^{-1}\dfrac{2M\sqrt{\dfrac{q^2B_0^2\sigma}{4M}+\omega_0^2(M\sigma-1)e^{2\Gamma\,t}}}{qB_0}-tanh^{-1}\dfrac{2M\sqrt{\dfrac{q^2B_0^2\sigma}{4M}+\omega_0^2(M\sigma-1)}}{qB_0} \right]\nonumber\\
&-\dfrac{(n+l)\sqrt{(\Delta-M\omega_0^2)\sigma}}{\Gamma}\left[tanh^{-1}\sqrt{\dfrac{q^2B_0^2\sigma+4M\omega_0^2(M\sigma-1)e^{2\Gamma\,t}}{q^2B_0^2\sigma}}-tanh^{-1}\sqrt{\dfrac{q^2B_0^2\sigma+4M\omega_0^2(M\sigma-1)}{q^2B_0^2\sigma}} \right]\nonumber\\
&-\dfrac{(n+l)\omega_0}{\Gamma}\left[tan^{-1}\dfrac{2M\omega_0 e^{\Gamma\,t}}{qB_0}-tan^{-1}\dfrac{2M\omega_0 }{qB_0}  \right]+\dfrac{(n+l)\sqrt{M\Delta-M^2\omega_0^2}}{2\Gamma\,M}\,log\,\dfrac{q^2B_0^2 e^{-2\Gamma\,t}+4M^2\omega_0^2}{q^2B_0^2 +4M^2\omega_0^2}~.
\end{align}

 
\noindent {\bf $\langle b \rangle $ Set-I~,~Case II}
\vskip .10cm
\noindent Here we set the constants 
\begin{eqnarray}
\vartheta=\Gamma~,~\delta=0~,\Lambda=\Gamma~.\label{case2}
\end{eqnarray}
So, the 
situation can be depicted by the following relations, 
\begin{eqnarray}
f(t)= e^{-\Gamma t}~;~\omega(t)={\omega_0}~;~B(t)=B_0\,e^{\Gamma\,t}~.
\label{10x}
\end{eqnarray}
Therefore, substituting  Eqn.(\ref{case2}) in Eqn(s).(\ref{expgenNC1},\ref{expgenNC2}), the reduced form of the NC parameters for this case are as follows,
\begin{align}
\theta(t)=\dfrac{8Me^{-\Gamma\,t}}{q^2B_0^2+4M^2\omega^2_0\,}
 \left[\sqrt{\dfrac{q^2B_0^2\sigma }{4M}+\omega_0^2\,\left(M\sigma\,-1\right)}-\dfrac{qB_0}{2M} \right]~,~                       
\Omega(t)=  -qB_0\,e^{\Gamma\,t}+2e^{\Gamma\,t}\sqrt{M\Delta\,-M^2\omega_0^2\,}                                                                                                          ~. \label{exp2conNC}
\end{align}
The point that is to be noted is that the multiplication of the two time dependent NC parameters obtained for this case reduces to a constant value and later we observe that the constant value is equal to the same found for another case discussed in Eqn.(\ref{-2ratdecNC}).~It can be checked that in the limit $B\rightarrow0$, the expressions for $\theta(t)$ and $\Omega(t)$ reduce 
to those obtained in \cite{SG}.~Substituting these relations in the expression for $c(t)$ in Eqn.(\ref{c}), we 
get,
\begin{eqnarray}
c(t)&=&\dfrac{1}{q^2B_0^2+4M^2\omega_0^2} \left[\left(4M^2\omega_0^2+2qB_0\sqrt{M\Delta-M^2\omega_0^2} \right)\sqrt{\dfrac{q^2B_0^2\sigma  }{4M}+\omega_0^2(M\sigma-1)}         \right.\nonumber \\
&&\left.          -2qB_0M\omega_0^2-\dfrac{q^2B_0^2}{M}\sqrt{M\Delta-M^2\omega_0^2}              \right]+\sqrt{\dfrac{\Delta}{M}-\omega_0^2}~.                                                                                \label{exp2conC}
\end{eqnarray}
Substituting the expressions of $a(t)\,,\rho(t)$\, and $c(t)$\, in Eqn.(\ref{eqn31})\,,\,we can get an expression for the phase in a closed form in the following way.\\ 
\begin{align}
\Theta_{\,n\,,\,l}(t)\,=\dfrac{(n\,+\,l)}{q^2B_0^2+4M^2\omega_0^2} &\left[\left(4M^2\omega_0^2+2qB_0\sqrt{M\Delta-M^2\omega_0^2} \right)\sqrt{\dfrac{q^2B_0^2\sigma }{4M}+\omega_0^2(M\sigma-1)}         \right.\nonumber \\
&\left.          -2qB_0M\omega_0^2-\dfrac{q^2B_0^2}{M}\sqrt{M\Delta-M^2\omega_0^2}              \right]t+(n+l)\left[\sqrt{\dfrac{\Delta}{M}-\omega_0^2}-\dfrac{\sigma}{\mu^2}\right]t~.
\end{align}
In this case the phase is varying linearly with respect to time. In the limit $B_0\rightarrow\,0$ ,~we can easily recover the same form of the phase factor corresponding to the solution set Ib obtained in \cite{SG}.\\
{\bf $\langle c \rangle $ Set-I~,~ Case III }
\vskip .20cm
Here we set the constants 
\begin{eqnarray}
\vartheta=\Gamma~,~\delta=0~,~\Lambda=-\Gamma~.              \label{case3}
\end{eqnarray}
 So, the 
situation can be depicted by the following relations, 
\begin{eqnarray}
f(t)= e^{-\Gamma\,t}~,~\omega(t)={\omega_0}~,~B(t)=B_0\,e^{-\Gamma\,t}~.
\label{10x}
\end{eqnarray}
Therefore, substituting Eqn.(\ref{case3}) in Eqn(s).(\ref{expgenNC1},\ref{expgenNC2}), the reduced form of the NC parameters for this case are as follows,
\begin{align}
\theta(t)=\dfrac{8Me^{-\Gamma\,t}}{q^2B_0^2e^{-4\Gamma\,t}+4M^2\omega^2_0}
 &\left[\sqrt{\dfrac{q^2B_0^2\sigma\,e^{-4\Gamma\,t}}{4M}+\omega_0^2(M\sigma-1)}-\dfrac{qB_0e^{-2\Gamma\,t}}{2M} \right]                    
  \nonumber\\
\Omega(t)&=-qB_0e^{-\Gamma\,t}+2e^{\Gamma\,t}\sqrt{M\Delta-M^2\omega_0^2}~.                              \label{exp2decNC}
\end{align}
Substituting these relations in the expression for $c(t)$ in Eqn.(\ref{c}), we 
get,
\begin{eqnarray}
c(t)&=&\dfrac{1}{q^2B_0^2e^{-4\Gamma\,t}+4M^2\omega_0^2} \left[\left(4M^2\omega_0^2+2qB_0e^{-2\Gamma\,t}\sqrt{M\Delta-M^2\omega_0^2} \right)\sqrt{\dfrac{q^2B_0^2\sigma e^{-4\Gamma\,t}}{4M}+\omega_0^2(M\sigma-1)}         \right.\nonumber \\
&&\left.          -2qB_0M\omega_0^2e^{-2\Gamma\,t}-\dfrac{q^2B_0^2e^{-4\Gamma\,t}}{M}\sqrt{M\Delta-M^2\omega_0^2}              \right]+\sqrt{\dfrac{\Delta}{M}-\omega_0^2} ~.
\label{exp2decC}
\end{eqnarray}
Substituting the expressions of $a(t)\,,\rho(t)$\, and $c(t)$\, in Eqn.(\ref{eqn31})\,,\,we can get an expression for the phase factor in a closed form as ,
\begin{align}
\small
\Theta_{\,n\,,\,l}(t)\,&=(n+l)\int_0^t \left[c(t)-\dfrac{a}{\rho^2}\right]~dT=(n+l)\left[\sqrt{\dfrac{\Delta}{M}-\omega_0^2}-\dfrac{\sigma}{\mu^2}\right]t\nonumber\\&+\dfrac{(n+l)\omega_0\sqrt{(M\sigma-1)}}{2\Gamma}
\,log\,\dfrac{\omega_0\sqrt{M\sigma-1}\,e^{2\Gamma\,t}+\sqrt{\dfrac{q^2B_0^2\sigma}{4M}+\omega_0^2(M\sigma-1)e^{4\Gamma\,t}} }{\omega_0\sqrt{M\sigma-1}+\sqrt{\dfrac{q^2B_0^2\sigma}{4M}+\omega_0^2(M\sigma-1)}}
\nonumber\\
&+\dfrac{(n+l)\omega_0}{2\Gamma}
\left[ tan^{-1}\dfrac{\omega_0 e^{2\Gamma\,t}}{\sqrt{\dfrac{q^2B_0^2\sigma}{4M}+\omega_0^2(M\sigma-1)e^{4\Gamma\,t}}}-
tan^{-1}\dfrac{\omega_0 }{\sqrt{\dfrac{q^2B_0^2\sigma}{4M}+\omega_0^2(M\sigma-1)}}
\right]\nonumber\\
&+\dfrac{(n+l)\sqrt{M\Delta-M^2\omega_0^2}}{2M\Gamma} \left[tanh^{-1}\dfrac{2M\sqrt{\dfrac{q^2B_0^2\sigma}{4M}+\omega_0^2(M\sigma-1)e^{4\Gamma\,t}}}{qB_0}-tanh^{-1}\dfrac{2M\sqrt{\dfrac{q^2B_0^2\sigma}{4M}+\omega_0^2(M\sigma-1)}}{qB_0} \right]\nonumber\\
&-\dfrac{(n+l)\sqrt{(\Delta-M\omega_0^2)\sigma}}{2\Gamma}\left[tanh^{-1}\sqrt{\dfrac{q^2B_0^2\sigma+4M\omega_0^2(M\sigma-1)e^{4\Gamma\,t}}{q^2B_0^2\sigma}}-tanh^{-1}\sqrt{\dfrac{q^2B_0^2\sigma+4M\omega_0^2(M\sigma-1)}{q^2B_0^2\sigma}} \right]\nonumber\\
&-\dfrac{(n+l)\omega_0}{2\Gamma}\left[tan^{-1}\dfrac{2M\omega_0 e^{2\Gamma\,t}}{qB_0}-tan^{-1}\dfrac{2M\omega_0 }{qB_0}  \right]+\dfrac{(n+l)\sqrt{M\Delta-M^2\omega_0^2}}{4\Gamma\,M}\,log\,\dfrac{q^2B_0^2 e^{-4\Gamma\,t}+4M^2\omega_0^2}{q^2B_0^2 +4M^2\omega_0^2}~.
\end{align}
\vskip .20cm

{\bf $\langle d \rangle $ Set-I~,~Case IV }
\vskip .20cm
\vskip .20cm
Here we set the constants 
\begin{eqnarray}
\vartheta=\delta=\Lambda=\Gamma~.~\label{case4}
\end{eqnarray}
 So, the 
situation can be depicted by the following relations, 
\begin{eqnarray}
f(t)= e^{-\Gamma\,t}~;~\omega(t)={\omega_0}e^{-\Gamma\,t/2}~;~B(t)=B_0\,e^{\Gamma\,t}~.
\end{eqnarray}
Substituting  Eqn.(\ref{case4}) in Eqn(s).(\ref{expgenNC1},\ref{expgenNC2}),~the reduced form of the NC parameters for this case are as follows,

\begin{align}
\theta(t)= \dfrac{8Me^{-\Gamma\,t}}{q^2B_0^2+4M^2\omega_0^2e^{-\Gamma\,t}}\left[\sqrt{\dfrac{q^2B_0^2\sigma}{4M}+\omega_0^2e^{-\Gamma\,t}(M\sigma-1)}-\dfrac{qB_0}{2M} \right]\,,\,          
\Omega(t)=-qB_0e^{\Gamma\,t}+2e^{\Gamma\,t}\sqrt{M\Delta-M^2\omega_0^2e^{-\Gamma\,t}}~.\label{exp3incNC}
\end{align}
Substituting these relations in the expression for $c(t)$ in Eqn.(\ref{c}), we 
get,
\begin{eqnarray}
c(t)&=&\dfrac{1}{q^2B_0^2+4M^2\omega_0^2e^{-\Gamma\,t}} \left[\left(4M^2\omega_0^2e^{-\Gamma\,t}+2qB_0\sqrt{M\Delta-M^2\omega_0^2e^{-\Gamma\,t}} \right)\sqrt{\dfrac{q^2B_0^2\sigma}{4M}+\omega_0^2e^{-\Gamma\,t}(M\sigma-1)}         \right.\nonumber \\
&&\left.          -2qB_0M\omega_0^2e^{-\Gamma\,t}-\dfrac{q^2B_0^2}{M}\sqrt{M\Delta-M^2\omega_0^2e^{-\Gamma\,t}}              \right]+\sqrt{\dfrac{\Delta}{M}-\omega_0^2e^{-\Gamma\,t}}~.\label{exp3incC}
\end{eqnarray}
We are able to obtain the exact form of the phase factor and it has been shown in the Appendix.

\subsection{Solution Set-II for Ermakov-Pinney equation: Rationally decaying solutions}
\subsubsection{The Solution Set}
We now consider rationally decaying solutions of the EP equation similar to that used in~\cite{Dey} which is of the form,
\begin{eqnarray}
a(t)=\dfrac{\sigma\,\left(1+\dfrac{2}{k}\right)^{\,(k+2)/k}}{(\Gamma{t}+\chi)^{\,(k+2)/k}}~,~
b(t)=\dfrac{\Delta\,\left(\dfrac{k}{k+2} \right)^{(2-k)/k} }{(\Gamma{t}+\chi)^{\,(k-2)/k}}     ~,~\rho(t)=\dfrac{\mu\left(1+\dfrac{2}{k}\right)^{1/k} }{(\Gamma{t}+\chi)^{1/k}}~;
\label{EPsoln2} 
\end{eqnarray}
where $\sigma$, $\Delta$, $\mu$, $\Gamma$ and $\chi$ are constants such that $(\Gamma{t}+\chi)~\neq~0$, and $k$ is an integer.~Substituting the expressions of $a(t)$, $b(t)$, and $\rho(t)$ in the EP equation, we can easily verify the relation between these constants to be as follows, 
\begin{equation}
\Gamma^2\mu=(k+2)^2\,(\sigma\Delta\mu-\frac{\xi^2\sigma^2}{\mu^3}).
\label{EPreln2}
\end{equation}



\subsubsection{Study of the corresponding eigenfunctions}
The eigenfunction of the invariant operator $I(t)$ [given by 
Eqn.(\ref{eqn28})] for this solution Set-II is of the following form,
\begin{eqnarray}
\phi_{n\,,\,m-n}(r,\theta)=\lambda_{n}\,\dfrac{{(i\mu)}^{\,m}}{\sqrt{m!}}\left[\dfrac{k+2}{k(\Gamma{t}+\chi)}\right]^{m/k}    r^{n-m}e^{i(m-n)\theta-\dfrac{[\sigma\,(k+2)\,+\,i\mu^2\Gamma]\,\,(\Gamma{t}+\chi)^{2/k}\,\,\,k^{2/k} }{2\sigma\,(k+2)^{\,(k+2)/k}\mu^2}r^2}\nonumber \\
\times\,\,\,U\left(-m,1-m+n,\,\dfrac{r^2[k(\Gamma{t}+\chi)]^{2/k}}{\mu^2\left(k+2\right)^{2/k}}\,\right)
\label{eqn51}
\end{eqnarray}
where $\lambda_n$ is given by 
\begin{eqnarray}
\lambda_n^{\,2}=\dfrac{1}{\pi\,n!\mu^{2n+2}}\left[\dfrac{k(\Gamma{t}+\chi)}{k+2}\right]^{2(1+n)/k}.
\label{eqn51lam}
\end{eqnarray}
In order to get the eigenfunction of the Hamiltonian $H(t)$,~we need to calculate the associated phase factor.~Once again for this we need to fix up the forms of the damping factor $f(t)$,~angular frequency $\omega(t)$ of the oscillator and the applied magnetic field $B(t)$.~In order to 
explore the solution of $H(t)$ for rationally decaying coefficients, we choose a rationally decaying form for $\omega(t)$,~$B(t)$
and set $f(t)=1$.~Thus,~we have the following relations, 
\begin{eqnarray}
\eta(t)=0\,\,\Rightarrow\,\,f(t)=1~,~
\omega(t)=\dfrac{\omega_0}{(\Gamma\,t+\chi)}~,~B(t)=\dfrac{B_0}{(\Gamma\,t+\chi)}.
\label{2ratdec}
\end{eqnarray}

{\bf $\langle e \rangle $ Set-II~,~Case I }\vskip .20cm

\noindent As we want to study how the nature of the rationally decaying solution gets altered when the system is placed in a 
magnetic field, we set $k=2$ in Eqn.(\ref{EPsoln2}). The system has been studied without applying any external field 
for this particular $k$ parameter in an 
earlier communication \cite{SG}.  

\noindent When we set $k=2$,the set $a(t)$, $b(t)$ and $\rho(t)$ takes the following simplified form,
\begin{eqnarray}
a(t)=\dfrac{4\sigma}{(\Gamma{t}+\chi)^{\,2}}\,\,,\,\,b(t)\,=\,\Delta\,\,,\,\,\rho(t)=\left[\dfrac{2\mu^{\,2}}{\Gamma{t}+\chi}\right]^{1/2}.\label{2rat}
\end{eqnarray}
Substituting the expressions for $a(t)$, $b(t)$, $\omega(t)$ , $f(t)$ and $B(t)$ in the 
Eqn(s).(\ref{a}, \ref{b}), we get the time dependent NC parameters as,
\begin{align}
\theta(t)=\dfrac{8M}{q^2B_0^2+4M^2\omega^2_0}&\left[\sqrt{ \dfrac{q^2B_0^2\sigma}{M}+\omega_0^2 4\sigma M-\omega_0^2(\Gamma\,t+\chi)^2}-\dfrac{qB_0}{2M}(\Gamma\,t+\chi)  \right]                      \nonumber \\
&\Omega(t)=- \dfrac{qB_0}{(\Gamma\,t+\chi)}+2\sqrt{M\Delta-\dfrac{M^2\omega_0^2}{(\Gamma\,t+\chi)^2}}     ~.                            \label{2ratdecNC}
\end{align}
Substituting these relations in the expression for $c(t)$ in Eqn.(\ref{c}) gives, 
\begin{eqnarray}
c(t)&=&\dfrac{1}{4M^2\omega^2_0+q^2B_0^2}\left[\left( \dfrac{4M^2\omega_0^2}{(\Gamma\,t+\chi)^2}+\dfrac{2qB_0\sqrt{M\Delta(\Gamma\,t+\chi)^2-M^2\omega_0^2}}{(\Gamma\,t+\chi)^2}\right)
\sqrt{ \dfrac{q^2B_0^2\sigma}{M}-\omega_0^2 (\Gamma\,t+\chi)^2 +4\omega_0^2\sigma M   } \right.\nonumber\\ 
&&\left.-\dfrac{2qB_0M\omega_0^2}{\Gamma\,t+\chi}-\dfrac{q^2B_0^2\sqrt{M\Delta(\Gamma\,t+\chi)^2-M^2\omega_0^2}}{M(\Gamma\,t+\chi)} \right]+\sqrt{\dfrac{\Delta}{M}-\dfrac{\omega_0^2}{(\Gamma\,t+\chi)^2}}~.                              \label{2ratdecC}
\end{eqnarray}
The additional terms that appear in the expression due to the presence of the magnetic field are mostly seen to be 
decaying functions of time. Their contribution becomes more evident when we study the evolution of expectation 
value of energy with time in a later section. 
We are able to obtain the exact form of the phase factor and it has been shown in the Appendix.

 \vskip .15cm

{\bf $\langle f \rangle $ Set-II~,~Case II }\vskip .20cm

\noindent It is observed from the solution set given by Eqn.(\ref{EPsoln2}), that the time dependent parameters 
$a(t)$ and $\rho(t)$ vanish, while $b(t)$ diverges if we set $k=-2$. In order to avoid this and study the system at this critical value of $k$, we choose the corresponding solution set to be,
\begin{eqnarray}
a(t)=\dfrac{\sigma}{(\Gamma{t}+\chi)^{\,(k+2)/k}}~,~
b(t)=\dfrac{\Delta}{(\Gamma{t}+\chi)^{\,(k-2)/k}}     ~,~\rho(t)=\dfrac{\mu}{(\Gamma{t}+\chi)^{1/k}}.
\label{k2} 
\end{eqnarray}
We now set $k=-2$ in Eqn. (\ref{k2}) to obtain the following set,
\begin{eqnarray}
a=\sigma ~~,~~ b(t)=\dfrac{\Delta}{(\Gamma\,t+\chi)^2}~~,~~ \rho=\mu \sqrt{\Gamma\,t+\chi}\,.
\label{ratsol}
\end{eqnarray}
Substituting these relations in the EP equation 
the constraint condition is found to be,
\begin{equation}
-\mu^4\Gamma^2+4\sigma\Delta\mu^4=4\,\xi^2\sigma^2\,.\label{rat2const}
\end{equation} 
The above relation matches with that found in Eqn.(\ref{EPreln1}) while considering $\vartheta=\Gamma$.   

As the  Eqn.({\ref{eqn51}}) vanish at $k=-2$, the eigenfunction of the invariant for this solution is needed to be calculated separately and it is of the following form,
\begin{eqnarray}
\phi_{n\,,\,m-n}(r,\theta)=\lambda_{n}\,\dfrac{{(i\mu\sqrt{(\Gamma\,t+\chi})}^{\,m}}{\sqrt{m!}}    r^{n-m}e^{i(m-n)\theta-\dfrac{2\sigma-i\mu^2\Gamma }{4\sigma\mu^2(\Gamma\,t+\chi)}r^2}
~U\left(-m,1-m+n,\,\dfrac{r^2}{\mu^2(\Gamma\,t+\chi)}\,\right)
\label{eqn511}
\end{eqnarray}
where $\lambda_n$ is given by
\begin{eqnarray}
\lambda_n^{\,2}=\dfrac{1}{\pi\,n![\mu^2(\Gamma\,t+\chi)]^{1+n}}.
\label{eqn51lam1}
\end{eqnarray}

Again we consider the same explicit forms of the damping factor $f(t)$, angular frequency 
$\omega(t)$ and applied 
magnetic field $B(t)$ as in Eqn.(\ref{2ratdec}),
\begin{eqnarray}
f(t)=1~,~ \omega(t)=\dfrac{\omega_0}{(\Gamma\,t+\chi)}~,~ B(t)=\dfrac{B_0}{(\Gamma\,t+\chi)}~. \label{2ratinc}
\end{eqnarray}
Substituting the expressions for $a(t)$, $b(t)$, $\omega(t)$ , $f(t)$ and $B(t)$ in the 
Eqn(s).(\ref{a}, \ref{b}), we get the time dependent NC parameters as,
\begin{align}
\theta(t)= \dfrac{8M(\Gamma\,t+\chi)}{q^2B_0^2+4M^2\omega_0^2}\left[\sqrt{\dfrac{q^2B_0^2\sigma}{4M}+\omega_0^2(M\sigma-1)}-\dfrac{qB_0}{2M} \right]                  ~,~ \Omega(t)= 2\dfrac{\sqrt{M\Delta-M^2\omega_0^2}}{(\Gamma\,t+\chi)}-\dfrac{qB_0}{(\Gamma\,t+\chi)}~.                                                                                 \label{-2ratdecNC}
\end{align}
Once again it is interesting to note that a constant value is found after multiplication of the two time dependent NC parameters obtained above. Here we recall Eqn.(\ref{exp2conNC}) where we discussed that the multiplication of two time dependent NC parameters reduces to a constant value which is equal to the same obtained for this case.  

\noindent Substituting these relations in the expression for $c(t)$ in Eqn.(\ref{c}) gives, 
\begin{eqnarray}
c(t)&=&\dfrac{1}{(4M^2\omega^2_0+q^2B_0^2)(\Gamma\,t+\chi)}\left[\left( 4M^2\omega_0^2+2qB_0\sqrt{M\Delta-M^2\omega_0^2}\right)
\sqrt{ \dfrac{q^2B_0^2\sigma}{4M}+\omega_0^2(M\sigma-1)  } \right.\nonumber\\ 
&&\left.-2qB_0M\omega_0^2-\dfrac{q^2B_0^2\sqrt{M\Delta-M^2\omega_0^2}}{M} \right]+\dfrac{1}{(\Gamma\,t+\chi)}\sqrt{\dfrac{\Delta}{M}-\omega_0^2 }~.
 \label{-2ratdecC}
\end{eqnarray}
Substituting these expressions for $a(t)$, $\rho(t)$ and $c(t)$ 
in Eqn.(\ref{eqn31}), we get the following expression 
for the phase factor in a closed form as,
\begin{eqnarray}
\Theta_{\,n, l\,}(t)&=&\dfrac{(n+l)}{(4M^2\omega^2_0+q^2B_0^2)\Gamma}\left[\left( 4M^2\omega_0^2+2qB_0\sqrt{M\Delta-M^2\omega_0^2}\right)
\sqrt{ \dfrac{q^2B_0^2\sigma}{4M}+\omega_0^2(M\sigma-1)  } \right.\nonumber\\ 
&&\left.-2qB_0M\omega_0^2-\dfrac{q^2B_0^2\sqrt{M\Delta-M^2\omega_0^2}}{M} \right]ln\dfrac{(\Gamma\,t+\chi)}{\chi}+\dfrac{(n+l)}{\Gamma}\left[\sqrt{\dfrac{\Delta}{M}-\omega_0^2 }-\dfrac{\sigma}{\mu^2}\right]~ln\dfrac{(\Gamma\,t+\chi)}{\chi}~.\nonumber\\
\end{eqnarray}



\section{Analysis of the expectation value of energy}
In this section, we intend to calculate the expectation value of energy. It is 
shown in \cite{SG} that the 
expectation value of energy $\braket{E_{n,m-n}(t)}$ with 
respect to energy eigenstate $\psi_{n,m-n}(r,\theta,t)$ can be expressed as,

\begin{align}
\braket{E_{n,m-n}(t)}&=\dfrac{1}{2}\,(n+m+1)\left[b(t)\rho^2(t)+\dfrac{a(t)}{\rho^2(t)}+\dfrac{\dot{\rho}^2(t)}{a(t)} \right]+c(t)\,(n-m)\,\,.
\end{align}
Substituting the expression of $c(t)$ in the above equation,~the 
expectation value of energy for our model takes the following form,
\begin{align}
\braket{E_{n,m-n}(t)}&=\dfrac{1}{2}\,(n+m+1)\left[b(t)\rho^2(t)+\dfrac{a(t)}{\rho^2(t)}+\dfrac{\dot{\rho}^2(t)}{a(t)} \right]\nonumber \\
&+\dfrac{(n-m)}{2}\left[\dfrac{qB(t)f(t)}{M} \left( 1+\dfrac{\theta(t)\Omega(t)}{4}\right) +\dfrac{\Omega(t)f(t)}{M}+  \left(\dfrac{q^2B^2(t)f(t)}{4M}+\dfrac{M\omega^2(t)}{f(t)}  \right)\theta(t) \right]~;\label{Energy}
\end{align}
which reduces to the same obtained in \cite{SG} in the limit $B\rightarrow\,0$.\\The energy expression depends on the charge explicitly.~It contains both terms having linear and quadratic dependence on charge. So the energy does not remain invariant when the charge of the particle changes its sign.
Another notable point is that even when the frequency of oscillation $\omega{\rightarrow}0$,~and the applied field $B\rightarrow\,0$;~the 
expectation value of energy is non-zero. This is because all the three parameters of the Hamiltonian $a(t)$, $b(t)$ and $c(t)$ are finite even as $\omega{\rightarrow}0, B{\rightarrow}0$, as is clear from the Eqn(s).(\ref{a},\ref{b},\ref{c}). Now we will proceed to study the time-dependent behaviour of $\braket{E_{n,m-n}(t)}$ for various types of damping and applied magnetic field.

   

\subsection{Solution Set-I: Exponentially decaying solution}
For the exponentially decaying solution given by Eqn.(\ref{EPsoln1}), the energy expectation value takes the following form,
\begin{equation}
\braket{E_{n,m-n}(t)}=(n+m+1)\mu^2\Delta+c(t)\,(n-m)
\label{eqn90}
\end{equation}
where we have set the constant $\xi^2$ to unity and used the constraint relation given by Eqn.(\ref{EPreln1}).

\vskip 0.1cm



\begin{figure}[t]
\centering
\includegraphics[scale=0.4]{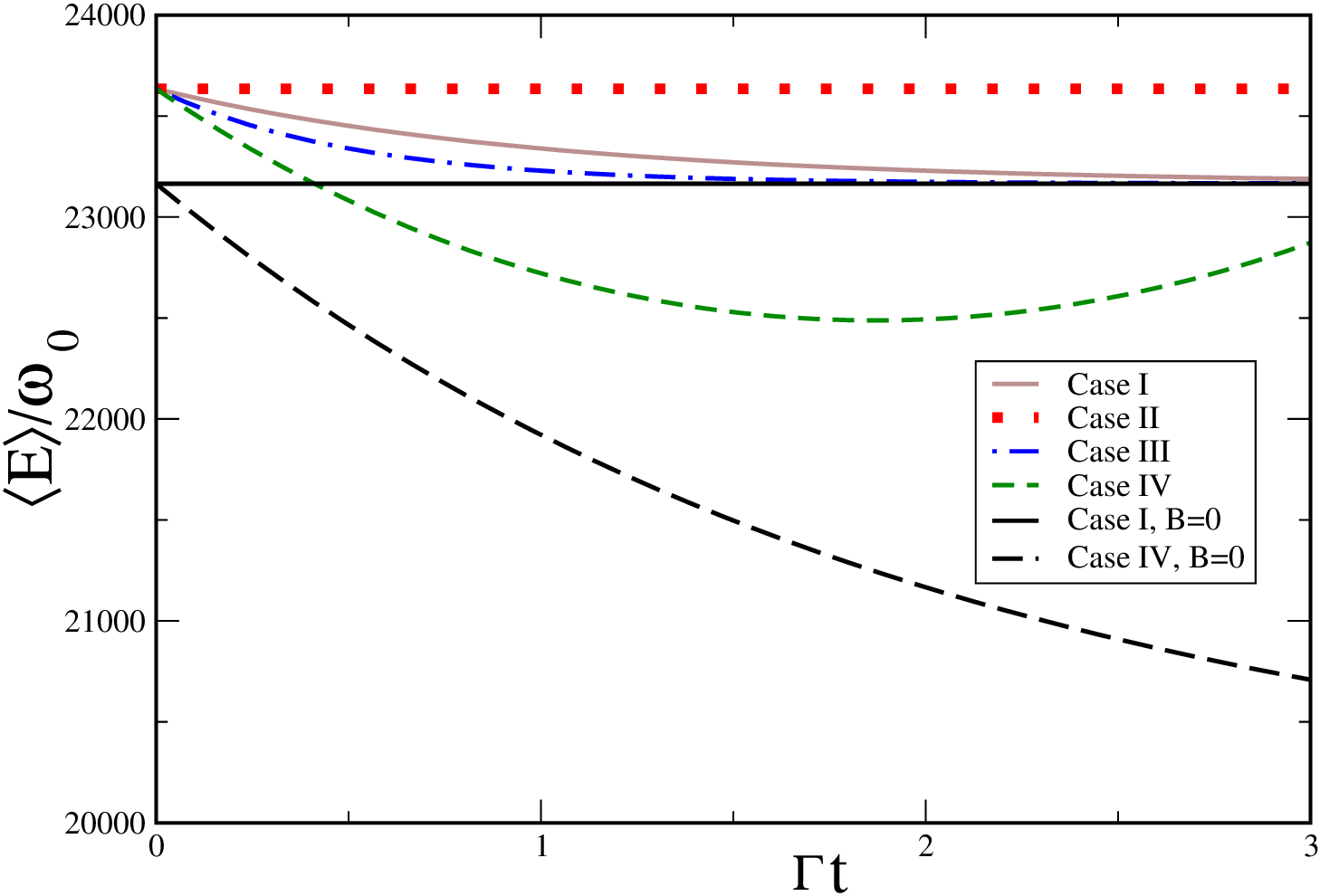}
\caption{\textit{A study of the variation of expectation value of energy, scaled by 
$\frac{1}{\omega_0}$ ($\frac{\langle E \rangle}{\omega_0}$) in order to make 
it dimensionless, as we vary $\Gamma$t (again a dimensionless quantity). Here 
we consider mass M=1, charge q=1, magnetic field $B_0$=$10^2$, $\mu$=1,$\Delta$=$10^7$, $\sigma$=$10^7$, 
$\omega_0$=$10^3$ and $\Gamma$=1 in natural units. The constants n=1 and m=0. 
The expectation value of 
energy $\langle E \rangle$ is  calculated for exponentially decaying solution set when Case I: $f(t)=e^{-\Gamma\,t}$ , $\omega(t)=\omega_0$ and  $B(t)=B_0$; Case II: $f(t)=e^{-\Gamma\,t}$ , $\omega(t)=\omega_0$ and  $B(t)=B_0\,e^{\Gamma\,t}$; Case III: $f(t)=e^{-\Gamma\,t}$ , $\omega(t)=\omega_0$ and  $B(t)=B_0e^{-\Gamma\,t}$ and Case IV: $f(t)=e^{-\Gamma\,t}$ , $\omega(t)=\omega_0e^{-\Gamma\,t/2}$ and  $B(t)=B_0e^{\Gamma\,t}$. While 
for Case I and Case III the energy first decreases, then becomes constant with 
time, 
for Case II the energy remains constant as we vary time. 
For Case IV the behaviour of energy with time is seen to be 
extremely interesting. It first decreases and then increases with time. Along 
with these, we have also plotted what happens in the absence of magnetic field 
for comparison. When the angular frequency of oscillation is a constant 
(Case I, Case II and Case III), if the magnetic field is set to zero, then the 
energy of the oscillator is a constant with time. So, the magnetic field, even 
when it is a constant brings about time variation in energy for an 
exponentially damped oscillator 
having a constant frequency. However, when
the angular frequency is decaying exponentially too (Case IV), then even when 
B=0, the energy decays with time. Nevertheless, the variation of energy is 
remarkably different from that seen in the presence of the field for Case IV.}}  
\label{fig1}
\end{figure}

\noindent{{\bf{$\langle A\rangle$ Set-I \,,\,Case I}}}
\vskip .20cm
\noindent Here we set $f(t)=e^{-\Gamma\,t}$ , $\omega(t)=\omega_0$ and  $B(t)=B_0$. With this the energy expression for the ground state takes the form,

\begin{align}
\braket{E_{n,-n}(t)}&=(n+1)\mu^2\Delta+\dfrac{n}{q^2B_0^2e^{-2\Gamma\,t}+4M^2\omega_0^2}\left[ -2qB_0M\omega_0^2e^{-\Gamma\,t}-\dfrac{q^2B_0^2}{M}e^{-2\Gamma\,t}\sqrt{M\Delta-M^2\omega_0^2}           \right.\nonumber\\
& \left. +\left(4M^2\omega_0^2+2qB_0e^{-\Gamma\,t}\sqrt{M\Delta-M^2\omega_0^2} \right)\sqrt{\dfrac{q^2B_0^2\sigma e^{-2\Gamma\,t} }{4M}+\omega_0^2(M\sigma-1)} \right]+n\,\sqrt{\dfrac{\Delta}{M}-\omega_0^2} .\label{exp1EN1}     
\end{align}
The nature of this energy expectation value depends on the value of the constants. Specifically, the sign of the charge plays a crucial role for determining the nature of this expectation value. It is noteworthy that in both the limits $t\rightarrow\infty$ and $B\rightarrow\,0$ this energy expression gains the same constant value which is already found in \cite{SG} for a damped oscillator in time dependent NC space.~Apart from it,~an inclusion of constant magnetic field in the system considered in \cite{SG} also makes the Hamiltonian non-hermitian after a certain limit of time beyond which the energy becomes imaginary. The 
condition for getting the expectation value of energy to be real is as follows,

\begin{equation}
t~\leq~\dfrac{1}{2\Gamma}\,ln \dfrac{q^2B_0^2\sigma}{4M\omega_0^2(1-M\sigma)}~.\label{exp1tim1}
\end{equation}
It is interesting to note that the upper bound of time below which the energy 
expectation value remains real does not depend on the sign of charge of the 
oscillator, although the value of energy itself does. The nature of variation 
of energy expectation value with time is shown in Fig.\ref{fig1}. Though the 
energy shows an initial decrease, eventually it tends to be a constant over 
time. 



\noindent{{\bf{$\langle B\rangle$ Set-I \,,\,Case II}}}\\
\noindent Here we set $f(t)=e^{-\Gamma\,t}$ , $\omega(t)=\omega_0$ and  $B(t)=B_0\,e^{\Gamma\,t}$. With this the energy expression for the ground state takes the form,
\begin{align}
\braket{E_{n,-n}(t)}&=(n+1)\mu^2\Delta+ 
\dfrac{n}{q^2B_0^2+4M^2\omega_0^2} \left[\left(4M^2\omega_0^2+2qB_0\sqrt{M\Delta-M^2\omega_0^2} \right)\sqrt{\dfrac{q^2B_0^2\sigma }{4M}+\omega_0^2(M\sigma-1)}         \right.\nonumber \\
&\left.          -2qB_0M\omega_0^2-\dfrac{q^2B_0^2}{M}\sqrt{M\Delta-M^2\omega_0^2}              \right]+n\sqrt{\dfrac{\Delta}{M}-\omega_0^2}=constant~.\label{exp1EN2}
\end{align} 
As expected we observe from Fig. \ref{fig1}, the energy is a constant 
over time. In the limit $B\rightarrow\,0$,~the constant value of energy reduces to the same obtained in \cite{SG} for a damped oscillator in time dependent NC space. 
\vskip .20cm
{\bf{$\langle C\rangle$ Set-I \,,\,Case III}}\\
\noindent Here we set $f(t)=e^{-\Gamma\,t}$ , $\omega(t)=\omega_0$ and  $B(t)=B_0e^{-\Gamma\,t}$. With this the energy expression for the ground state takes the form,
\begin{align}
\braket{E_{n,-n}(t)}&=(n+1)\mu^2\Delta+\dfrac{n}{q^2B_0^2e^{-4\Gamma\,t}+4M^2\omega_0^2} \left[   -2qB_0M\omega_0^2e^{-2\Gamma\,t}-\dfrac{q^2B_0^2e^{-4\Gamma\,t}}{M}\sqrt{M\Delta-M^2\omega_0^2}       \right.\nonumber \\
&\left.    +\left(4M^2\omega_0^2+2qB_0e^{-2\Gamma\,t}\sqrt{M\Delta-M^2\omega_0^2} \right)\sqrt{\dfrac{q^2B_0^2\sigma e^{-4\Gamma\,t}}{4M}+\omega_0^2(M\sigma-1)}                \right]+n~\sqrt{\dfrac{\Delta}{M}-\omega_0^2} ~. \label{exp1EN3}   
\end{align}
The nature of variation of the energy expectation value with time has almost 
the same characteristics as obtained in Case I. This is also observed 
graphically from Fig. \ref{fig1}. However, a closer observation tells us that 
in Case III the rate of decay is faster than in Case I for the same set of 
parameters. It must be because in Case III, unlike in Case I, the applied 
field is decaying as 
well with respect to time. The upper bound of time beyond which the system 
becomes non-physical (since the energy ceases to be real after this time) 
is also half of that obtained in Eqn.(\ref{exp1tim1}). Here the bound is 
found to be
\begin{equation}
t\leq\,\dfrac{1}{4\Gamma}ln \dfrac{q^2B_0^2\sigma}{4M\omega_0^2(1-M\sigma)}~.\label{exp1tim3}
\end{equation}
An important inference that can be drawn from the above study is that, in the 
presence of the external magnetic field, the energy of a damped oscillator 
decays with time if the field either decays with time or is atleast a 
constant. If the field tends to grow as fast as the damping factor decays, 
then the energy of the oscillator tends to be a constant with time. In 
Fig. \ref{fig1} we have also plotted the evolution of energy with time in 
this situation if the applied field is turned off. We see that if B=0, then 
the energy is a constant over time. This is expected from the energy 
expectation value expressions given by Eqns. [\ref{exp1EN1}, \ref{exp1EN2} and \ref{exp1EN3}]. Even a constant magnetic field is able to bring about time 
variation in this energy value.

\vskip .20cm
\noindent{{\bf{$\langle D\rangle$ Set-I, Case IV}}}\\
\noindent 
 Here we set $f(t)=e^{-\Gamma\,t}$ , $\omega(t)=\omega_0e^{-\Gamma\,t/2}$ and  $B(t)=B_0e^{\Gamma\,t}$. With this the energy expression for the ground state takes the form,
\begin{align}
\braket{E_{n,-n}(t)}&=(n+1)\mu^2\Delta+\dfrac{n}{q^2B_0^2+4M^2\omega_0^2e^{-\Gamma\,t}} \left[     -2qB_0 M\omega_0^2e^{-\Gamma\,t}-\dfrac{q^2B_0^2}{M}\sqrt{M\Delta-M^2\omega_0^2e^{-\Gamma\,t}}               \right.\nonumber \\
&\left.+\left(4M^2\omega_0^2e^{-\Gamma\,t}+2qB_0\sqrt{M\Delta-M^2\omega_0^2e^{-\Gamma\,t}} \right)\sqrt{\dfrac{q^2B_0^2\sigma}{4M}+\omega_0^2e^{-\Gamma\,t}(M\sigma-1)}  \right]    +n\sqrt{\dfrac{\Delta}{M}-\omega_0^2e^{-\Gamma\,t}}~. \label{exp2En}
\end{align}  
It can be verified that the above energy expectation value reduces to a decaying form as obtained in \cite{SG} in the absence of magnetic field. However, it 
is interesting to note that for certain choice of parameters (as shown in 
Fig. \ref{fig1}) the energy may also exhibit an initial growth with time. 
Infact for the given set of parameters in Fig. \ref{fig1}, it initially 
decays and then increases. Eventually, as $t~\rightarrow~\infty$, the energy 
tends to be a constant. In this situation, if we turn off the applied magnetic 
field, then we see from Fig \ref{fig1}, the energy decays off with time. 
However, due to the absence of the exponentially growing field, the energy of 
the oscillator is seen to be much lower. Also the exponential growth in energy 
seen in the presence of the field is absent within the time range studied.
The system also possess two lower bounds of time below which it becomes non-physical due to the imaginary value of energy.~The conditions are as follows,              
\begin{eqnarray}
t~ \geq~\dfrac{1}{\Gamma}ln\left[\dfrac{M\omega_0^2}{\Delta} \right]~;~
t~\geq~\dfrac{1}{\Gamma}ln\left[\dfrac{4M\omega_0^2(1-M\sigma)}{q^2B_0^2\sigma} \right]~.
\end{eqnarray}
The greater of the two bounds serves as the actual lower bound. 

\subsection{Solution Set-II: Rationally decaying solution}
In previous section we considered two different solution set generated from Eqn.(\ref{EPsoln2}).~The special form of the solution shown in Eqn.(\ref{2rat}) is directly produced by substituting $k=2$ in the Eqn.(\ref{EPsoln2}) and the special form of the solution shown in Eqn.(\ref{ratsol}) is obtained by substituting $k=-2$ in the modified form of Eqn.(\ref{EPsoln2}). 
\subsubsection{Set-II, Case I}
\begin{figure}[t]
\centering
\includegraphics[scale=0.4]{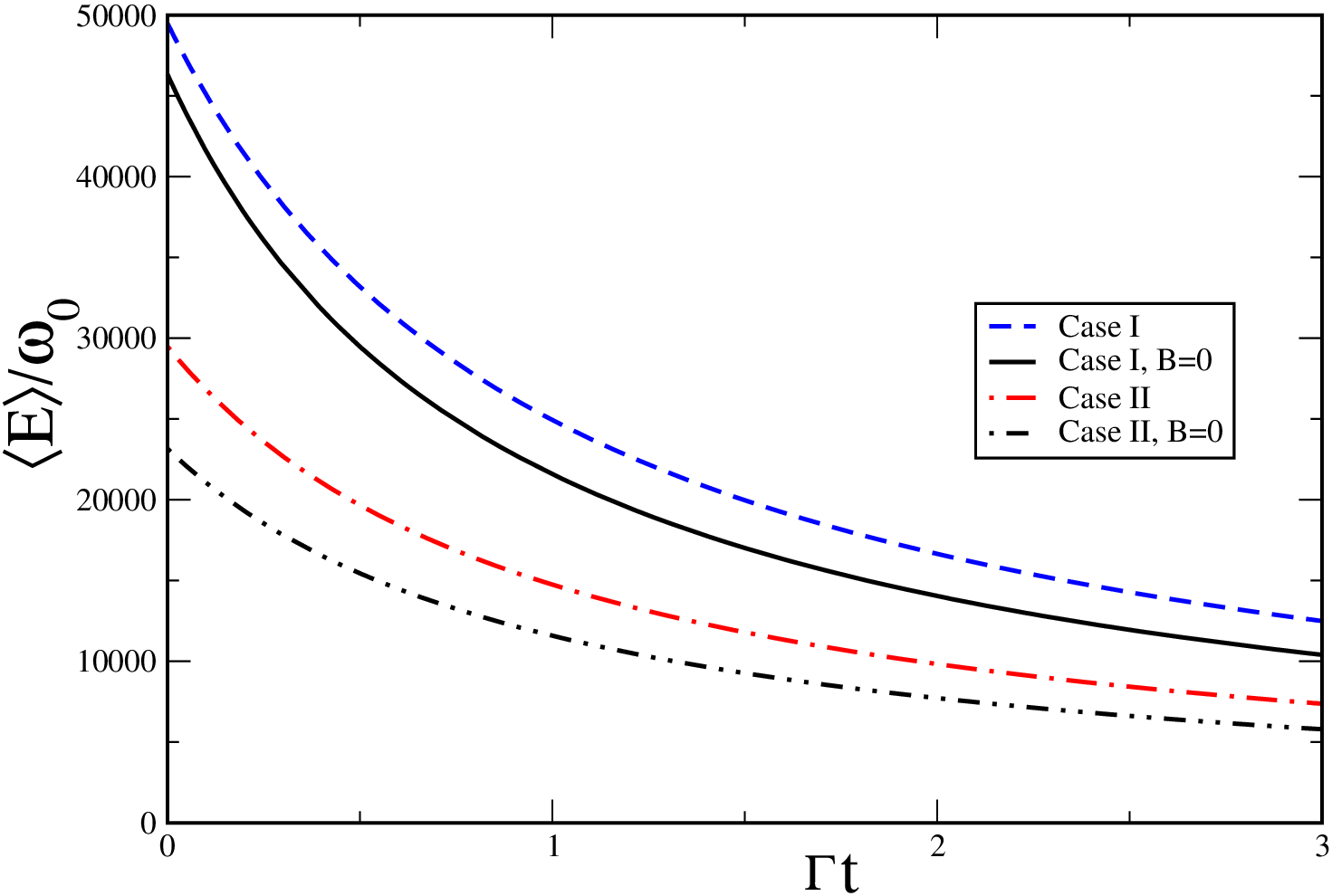}
\caption{\textit{A study of the variation of expectation value of energy, scaled by 
$\frac{1}{\omega_0}$ ($\frac{\langle E \rangle}{\omega_0}$) in order to make 
it dimensionless, as we vary $\Gamma$t (again a dimensionless quantity). Here 
we consider mass M=1, charge q=1, magnetic field $B_0$=$10^{20}$, $\mu$=1,$\Delta$=$10^7$, $\sigma$=$10^7$, 
$\omega_0$=$10^3$ and $\Gamma$=1 in natural units. The constants n=1 and m=0. 
The expectation value of 
energy $\langle E \rangle$ is calculated for rationally decaying 
solution set when $\langle A\rangle$ Case I: $a(t)=\dfrac{4\sigma}{(\Gamma{t}+\chi)^{\,2}}\,\,,\,\,b(t)\,=\,\Delta\,\,,\,\,\rho(t)=\left[\dfrac{2\mu^{\,2}}{\Gamma{t}+\chi}\right]^{1/2}$ and $\langle B\rangle$ Case II: $a=\sigma ~~,~~ b(t)=\dfrac{\Delta}{(\Gamma\,t+\chi)^2}~~,~~ \rho=\mu \sqrt{\Gamma\,t+\chi}$.  For both Case I and Case II the energy decays with time. However, the rate of 
decay is higher for Case I as compared to Case II. This is expected since two of the Hamiltonian 
parameters $a(t)$ and $\rho(t)$ are decaying with time for Case I, whereas only one parameter 
$b(t)$ is decaying while another one $\rho(t)$ is increasing with time for Case II. For comparison, we have also plotted Case I and II, in the absence of the magnetic field 
in the same plot. It is noteworthy, that although the energy 
of the oscillator decays even in the absence of the field, when we apply the 
field the energy is enhanced. This is due to the presence of magnetic energy 
in the system in this situation.}}  
\label{fig2}
\end{figure}

For the rationally decaying solution given by Eqn.(\ref{2rat}), the energy expectation value
takes the following form 
\begin{equation}
\braket{E_{n,m-n}(t)}=\dfrac{(n+m+1)}{2(\Gamma\,t+\chi)}\left[2\left(\dfrac{\sigma}{\mu^2} +\Delta\mu^2\right)+\dfrac{\mu^2\Gamma^2}{8\sigma} \right]+(n-m)c(t)~.\label{rat1EN} 
\end{equation}
where we have set the constant $\xi^2$ to unity and used the constraint relation given by Eqn.(\ref{EPreln2}).
Here we set $f(t)=1$ , $\omega(t)=\omega_0/(\Gamma\,t+\chi)$ and  $B(t)=B_0/(\Gamma\,t+\chi)$. With this the energy expression for the ground state takes the form,
\begin{align}
&\braket{E_{n,-n}(t)}=\dfrac{(n+1)}{2(\Gamma\,t+\chi)}\left[2\left(\dfrac{\sigma}{\mu^2} +\Delta\mu^2\right)+\dfrac{\mu^2\Gamma^2}{8\sigma} \right]
+\dfrac{n}{4M^2\omega^2_0+q^2B_0^2}\left[
-\dfrac{2qB_0M\omega_0^2}{\Gamma\,t+\chi}-\dfrac{q^2B_0^2\sqrt{M\Delta(\Gamma\,t+\chi)^2-M^2\omega_0^2}}{M(\Gamma\,t+\chi)}\right.
\nonumber\\
&\left.+\left( \dfrac{4M^2\omega_0^2}{(\Gamma\,t+\chi)^2}+\dfrac{2qB_0\sqrt{M\Delta(\Gamma\,t+\chi)^2-M^2\omega_0^2}}{(\Gamma\,t+\chi)^2}\right)
\sqrt{ \dfrac{q^2B_0^2\sigma}{M}-\omega_0^2 (\Gamma\,t+\chi)^2 +4\omega_0^2\sigma M   }  \right]+n\sqrt{\dfrac{\Delta}{M}-\dfrac{\omega_0^2}{(\Gamma\,t+\chi)^2}}\label{rat1EN2}~;
\end{align}
which is a decaying function of time and is seen to reduce to the same obtained in \cite{SG} in the limit $B\rightarrow\,0$.
The time range beyond which the system becomes non-physical due to imaginary energy expectation value is as follows,
\begin{equation}
\dfrac{1}{\Gamma}\left(\omega_0\sqrt{\dfrac{M}{\Delta}}-\chi \right)\leq~t\leq~\dfrac{1}{\Gamma}\left[\sqrt{\dfrac{q^2B_0^2\sigma}{M\omega_0^2}+4\sigma M} -\chi\right]~.\label{rat1tim2}
\end{equation}
In Fig. \ref{fig2} we have done a comparative study of the time variation of 
energy in this situation, when the applied magnetic field is present and when 
it is turned off. We see from the Figure that in both these cases the energy 
decays with time. However the energy of the oscillator is higher in the 
presence of the field due to the presence of the magnetic energy in the 
system. 

\subsubsection{Set-II, Case II}

For the rationally decaying solution given by Eqn.(\ref{ratsol}), the energy expectation value takes the following form,

\begin{equation}
\braket{E_{n,m-n}(t)}= \dfrac{(m+n+1)\mu^2\Delta}{\Gamma\,t+\chi}+c(t)(n-m)~.\label{rat2EN}
\end{equation}
where we have set the constant $\xi^2$ to unity and used the constraint relation given by Eqn.(\ref{rat2const}).
Here we set $f(t)=1$ , $\omega(t)=\omega_0/(\Gamma\,t+\chi)$ and  $B(t)=B_0/(\Gamma\,t+\chi)$. With this the energy expression for the ground state takes the form,


\begin{align}
\braket{E_{n,-n}(t)}=\dfrac{(n+1)\mu^2\Delta}{\Gamma\,t+\chi}+\dfrac{n}{(4M^2\omega^2_0+q^2B_0^2)(\Gamma\,t+\chi)}
\left[-2qB_0M\omega_0^2-\dfrac{q^2B_0^2\sqrt{M\Delta-M^2\omega_0^2}}{M}
 \right.\nonumber\\ 
\left. +\left( 4M^2\omega_0^2+2qB_0\sqrt{M\Delta-M^2\omega_0^2}\right)
\sqrt{ \dfrac{q^2B_0^2\sigma}{4M}+\omega_0^2(M\sigma-1)  }
\right]+\dfrac{n}{(\Gamma\,t+\chi)}\sqrt{\dfrac{\Delta}{M}-\omega_0^2 }~.
\end{align}
As we see from Fig. \ref{fig2} the corresponding energy expectation value decays off with time. 
Infact, it approaches zero at large time. However, the rate of decay is lower than Case I. This 
is expected since in Case II, unlike in Case I, while one of the Hamiltonian parameters $b(t)$ is decreasing with 
time, another parameter $\rho(t)$ is growing with time. The resultant energy is decreasing with 
time since the rate at which $b(t)$ is decaying (rate $\sim~t^{-2}$) is higher than the rate at 
which $\rho(t)$ is growing (rate $\sim~t^{1/2}$). When we turn off the field 
in this situation, we see from Fig. \ref{fig2}, the energy still decays off 
due to the decaying angular frequency of the oscillator. But in this situation 
the energy of the oscillator is reduced due to the absence of magnetic energy 
in the system.


\section{Conclusion}

In conclusion, our primary objective through this study has been to investigate the effect that an external 
time-dependent magnetic 
field has on a two dimensional damped harmonic oscillator in noncommutative space. The behaviour of the system under 
the influence of this time varying field is seen to be dependent on the nature of the field. For this purpose we have first set up 
the Hamiltonian for our system in the presence of a general magnetic field in noncommutative space. Then we map 
this Hamiltonian in terms of commutative variables by using a shift of variables connecting the noncommutative and 
commutative space, known in the literature as Bopp-shift.  We have then obtained the exact solution of this time dependent system in the presence of an applied time varying magnetic field by using the well known Lewis invariant which in turn leads to a non-linear differential equation known as the Ermakov-Pinney equation. Then we make various choices of the parameters 
of the system and study the solutions depending on these choices as we tune the applied magnetic field. In this study 
we have primarily considered two different sets of parameters for our damped system, namely, exponentially decaying solutions and rationally decaying solutions. Interestingly, the solutions obtained make it possible to integrate the 
expression of the phase factor exactly thereby giving an exact solution for the eigenstates of the Hamiltonian. Then we compute the expectation value of the Hamiltonian. Expectedly, the expectation value of the energy varies with time. For the exponentially 
decaying system, the nature of the time dependent magnetic field crucially determines the nature of evolution of the energy with time. 
There is basically an interplay between the damping factor, applied time varying magnetic field and time dependent 
angular frequency of the harmonic oscillator in determining the time evolution. While an exponentially growing magnetic field is able to maintain the 
energy to a constant value inspite of damping, a constant or an exponentially decaying field makes the energy fall off 
faster with damping. Remarkably, the presence of an exponentially decaying frequency along with the damping factor 
makes the behaviour of the system under the influence of an exponentially growing field even more interesting. While 
initially the energy decays off with time due to the damping present in the system, later the energy starts growing 
under the influence of the growing field. For the rationally decaying situation, even when damping is not present, a rationally decaying 
magnetic field in combination with a rationally decaying angular frequency is able to eventually damp out the energy of the 
oscillator. While the decaying oscillation corresponding to the Case I of rational EP solution  cannot remain physical at the zero value of energy due to the existance of upper bound of time, the same corresponding to the Case II of rational EP solution is physically damped out to be zero with time. 
However, we observe that at a given instant of time, the expectation value of energy
is greater in the presence of the magnetic field than when the field is turned off.
This is because magnetic energy is absent in the system in the latter case.

\section*{Appendix: Explicit forms of some phases}
We discussed about the eigenfunctions corresponding to the exponentially decaying EP solution set in section $4.1$. We mentioned that the exact form of the phase factor to construct the eigenfunction of a damped oscillator having exponentially decaying frequency in the presence of a magnetic field increasing exponentially with respect to time in NC space can be found [Set I, Case IV]. The phase factor is as follows,
\begin{align}
&\Theta_{\,n\,,\,l}(t)\,=
\dfrac{2(n+l)}{\Gamma} \left[\sqrt{\dfrac{\Delta}{M}-\omega_0^2}-\sqrt{\dfrac{\Delta}{M}-\omega_0^2 e^{-\Gamma\,t}} \right] +\dfrac{2(n+l)}{\Gamma}\left \lbrace \sqrt{\dfrac{q^2B_0^2\sigma}{4M}+\omega_0^2(M\sigma-1)}-\sqrt{\dfrac{q^2B_0^2\sigma}{4M}+\omega_0^2(M\sigma-1)e^{-\Gamma\,t}} \right \rbrace\nonumber\\
&+ (n+l)\left[-\dfrac{\sigma}{\mu^2}\right]t+\dfrac{i(n+l)}{M\Gamma}\sqrt{q^2B_0^2+4M\Delta}\left [tan^{-1}\dfrac{\sqrt{q^2B_0^2+4M\Delta}}{2\sqrt{M^2\omega_0^2-M\Delta}}-tan^{-1}\dfrac{\sqrt{q^2B_0^2+4M\Delta}}{2\sqrt{M^2\omega_0^2e^{-\Gamma\,t}-M\Delta}}  \right]
\nonumber\\
&+\dfrac{qB_0(n+l)}{2M\Gamma}\,log\,\dfrac{{(4M^2\omega_0^2+q^2B_0^2) \left \lbrace 
\left(\sqrt{\omega_0^2(M\sigma-1)+\dfrac{q^2B_0^2\sigma}{4M}e^{\Gamma\,t}}+\dfrac{qB_0\sigma}{2}e^{\Gamma\,t/2}        \right)^2  +\omega_0^2(M\sigma-1)^2 \right \rbrace}}{(4M^2\omega_0^2+q^2B_0^2\,e^{\Gamma\,t})\left \lbrace 
\left(\sqrt{\omega_0^2(M\sigma-1)+\dfrac{q^2B_0^2\sigma}{4M}}+\dfrac{qB_0\sigma}{2}        \right)^2 +\omega_0^2(M\sigma-1)^2 \right \rbrace} \nonumber\\
&+\dfrac{(n+l)\sqrt{q^2B_0^2+4M\Delta}}{2M\Gamma}\,log\,\dfrac{\left[(4M^2\omega_0^2+q^2B_0^2)\left \lbrace q^2B_0^2\sigma M^2\omega_0^2+4M^3\sigma\Delta\omega_0^2-2M\omega_0^2q^2B_0^2-4\omega_0^2M^2\Delta+e^{\Gamma\,t}\left(\dfrac{q^4B_0^4\sigma}{4}+q^2B_0^2\Delta\right)  \atop +e^{\Gamma\,t}q^2B_0^2\sigma M\Delta-2qB_0\sqrt{(q^2B_0^2+4M\Delta)\left(\omega_0^2[M\sigma-1]+\dfrac{q^2B_0^2\sigma}{4M}e^{\Gamma\,t} \right)(M\Delta e^{\Gamma\,t}-M^2\omega_0^2)}              
  \right\rbrace\right] }{\left[(4M^2\omega_0^2+q^2B_0^2 e^{\Gamma\,t})\left \lbrace q^2B_0^2\sigma M^2\omega_0^2+4M^3\sigma\Delta\omega_0^2-2M\omega_0^2q^2B_0^2-4\omega_0^2M^2\Delta+\dfrac{q^4B_0^4\sigma}{4}+q^2B_0^2\Delta  \atop +q^2B_0^2\sigma M\Delta-2qB_0\sqrt{(q^2B_0^2+4M\Delta)\left(\omega_0^2[M\sigma-1]+\dfrac{q^2B_0^2\sigma}{4M} \right)(M\Delta -M^2\omega_0^2)}              
  \right\rbrace\right] }\nonumber\\
&+\dfrac{(n+l)\sqrt{\sigma\Delta}}{\Gamma}\,log\,\dfrac{\left [{\omega_0^2M^2\sigma\Delta-M\Delta\omega_0^2-\dfrac{q^2B_0^2\sigma M\omega_0^2}{4}+\dfrac{q^2B_0^2\sigma\Delta}{2}e^{\Gamma\,t}\atop + \sqrt{q^2B_0^2\sigma\Delta\left(M\Delta e^{\Gamma\,t}-M^2\omega_0^2\right)\left(\omega_0^2[M\sigma-1]+\dfrac{q^2B_0^2\sigma}{4M}e^{\Gamma\,t} \right)} }\right ]}{\left[{\omega_0^2M^2\sigma\Delta-M\Delta\omega_0^2-\dfrac{q^2B_0^2\sigma M\omega_0^2}{4}+\dfrac{q^2B_0^2\sigma\Delta}{2}\atop +\sqrt{q^2B_0^2\sigma\Delta\left(M\Delta -M^2\omega_0^2\right)\left(\omega_0^2[M\sigma-1]+\dfrac{q^2B_0^2\sigma}{4M} \right)}}\right] }+\dfrac{(n+l)qB_0}{2M\Gamma}log\,\dfrac{q^2B_0^2+4M^2\omega_0^2e^{-\Gamma\,t}}
{q^2B_0^2+4M^2\omega_0^2}\nonumber\\
&+\dfrac{(n+l)iqB_0\sqrt{M\sigma-1}}{2M\Gamma} \times \nonumber\\
 &\left[log\dfrac{{M\sigma\Delta-\Delta-\dfrac{q^2B_0^2\sigma }{4}-2M\omega_0^2(M\sigma-1)e^{-\Gamma\,t}-2i\sqrt{(M\sigma-1)\left(\omega_0^2[M\sigma-1]e^{-\Gamma\,t}+\dfrac{q^2B_0^2\sigma}{4M}\right)(M\Delta-M^2\omega_0^2e^{-\Gamma\,t})} }}{{M\sigma\Delta-\Delta-\dfrac{q^2B_0^2\sigma }{4}-2M\omega_0^2(M\sigma-1)-2i\sqrt{(M\sigma-1)\left(\omega_0^2[M\sigma-1]+\dfrac{q^2B_0^2\sigma}{4M}\right)(M\Delta-M^2\omega_0^2)}}}\right]\nonumber\\
\end{align}

We also discussed about the eigenfunctions corresponding to the rationally decaying EP solution set in section $4.2$. We mentioned that the exact form of the phase factor to construct the eigenfunction of a damped oscillator having rationally decaying frequency in the presence of a magnetic field decaying rationally with respect to time in NC space 
[Set II, Case I] can be found. The phase factor is as follows,
\begin{align}
&\Theta_{\,n\,,\,l}(t)\,
= -\dfrac{2(n+l)}{\Gamma}\left(\dfrac{\sigma}{\mu^2}+\dfrac{qB_0M\omega_0^2}{4M^2\omega_0^2+q^2B_0^2} \right)log\,\dfrac{\Gamma\,t+\chi}{\chi}+\dfrac{4(n+l)M^2\omega_0^2}{(q^2B_0^2+4M^2\omega_0^2)\Gamma}\left[ \sqrt{\left( \dfrac{q^2B_0^2\sigma}{M}+4\omega_0^2\sigma\,M\right)\dfrac{1}{\chi^2}-\omega_0^2}  \right.\nonumber\\
-&\left.\sqrt{\left( \dfrac{q^2B_0^2\sigma}{M}+4\omega_0^2\sigma\,M\right)\dfrac{1}{(\Gamma\,t+\chi)^2}-\omega_0^2}\,+\omega_0\left\lbrace tan^{-1}\dfrac{\sqrt{M}~\omega_0\chi}{\sqrt{\left(q^2B_0^2\sigma+4\omega_0^2\sigma\,M^2\right)-\omega_0^2\chi^2M}}\right.\right.\nonumber\\
&\left.\left.
-tan^{-1}\dfrac{\sqrt{M}~\omega_0(\Gamma\,t+\chi)}{\sqrt{\left(q^2B_0^2\sigma+4\omega_0^2\sigma\,M^2\right)-\omega_0^2M(\Gamma\,t+\chi)^2}}
 \right\rbrace \right]
+\dfrac{2(n+l)qB_0}{q^2B_0^2+4M^2\omega_0^2}
 \left[\dfrac{1}{\Gamma}\left \lbrace \sqrt{\left(\Delta-\dfrac{M\omega_0^2}{\chi^2}\right)\left({4M^2\omega_0^2\sigma+\atop q^2B_0^2\sigma-M\omega_0^2\chi^2} \right)}\right.\right.\nonumber\\
 -&\left.\left.\sqrt{\left(M\Delta-\dfrac{M^2\omega_0^2}{(\Gamma\,t+\chi)^2}\right)\left(\dfrac{q^2B_0^2\sigma}{M}+4\omega_0^2\sigma\,M-\omega_0^2(\Gamma\,t+\chi)^2 \right)} \right\rbrace -\dfrac{2iM\omega_0^2}{\Gamma}\left\lbrace EllipticE\left(isinh^{-1}\left[\dfrac{i\omega_0\sqrt{M}(\Gamma\,t+\chi)}{\sqrt{q^2B_0^2\sigma+4\omega_0^2\sigma M^2}}\right]\right.\right.\right.
 \nonumber\\
 ,&\left.\left.\left.\dfrac{q^2B_0^2\sigma\Delta+4\omega_0^2\sigma M^2\Delta}{M^2\omega_0^4}\right) -EllipticE\left(isinh^{-1}\left[\dfrac{i\omega_0\sqrt{M}\chi}{\sqrt{q^2B_0^2\sigma+4\omega_0^2\sigma M^2}}\right],\dfrac{q^2B_0^2\sigma\Delta+4\omega_0^2\sigma M^2\Delta}{M^2\omega_0^4}\right) \right\rbrace     \right.\nonumber\\
&-\left. \dfrac{i(q^2B_0^2\sigma\Delta+4\omega_0^2\sigma M^2\Delta-M^2\omega_0^4)}{\Gamma M\omega_0^2}\left\lbrace
EllipticF\left(isinh^{-1}\left[\dfrac{i\omega_0\sqrt{M}(\Gamma\,t+\chi)}{\sqrt{q^2B_0^2\sigma+4\omega_0^2\sigma M^2}}\right],\dfrac{q^2B_0^2\sigma\Delta+4\omega_0^2\sigma M^2\Delta}{M^2\omega_0^4}\right)\right.\right.\nonumber\\
-&\left.\left.EllipticF\left(isinh^{-1}\left[\dfrac{i\omega_0\sqrt{M}\chi}{\sqrt{q^2B_0^2\sigma+4\omega_0^2\sigma M^2}}\right],\dfrac{{q^2B_0^2\sigma\Delta+\atop 4\omega_0^2\sigma M^2\Delta}}{M^2\omega_0^4}\right)    \right\rbrace          \right] +\dfrac{(n+l)}{\Gamma}\left[{\sqrt{\dfrac{\Delta}{M}(\Gamma\,t+\chi)^2-\omega_0^2}-\sqrt{\dfrac{\Delta}{M}\chi^2-\omega_0^2}}\right.\nonumber\\
+&\left.\omega_0\,tan^{-1}\sqrt{\dfrac{M\omega_0^2}{\Delta(\Gamma\,t+\chi)^2-M\omega_0^2}} -\omega_0\,tan^{-1}\sqrt{\dfrac{M\omega_0^2}{\Delta\chi^2-M\omega_0^2}} \right]-\dfrac{(n+l)q^2B_0^2}{M(4M^2\omega_0^2+q^2B_0^2)\Gamma}\left[\sqrt{M\Delta(\Gamma\,t+\chi)^2-M^2\omega_0^2}\right.
\nonumber\\ -&\left.\sqrt{M\Delta\chi^2-M^2\omega_0^2} 
+M\omega_0\,\left(tan^{-1}\dfrac{M\omega_0}{\sqrt{M\Delta(\Gamma\,t+\chi)^2-M^2\omega_0^2}}-\,tan^{-1}\dfrac{M\omega_0}{\sqrt{M\Delta\chi^2-M^2\omega_0^2}}   \right)\right]~.
\end{align}
Here $EllipticF$ and $EllipticE$ are the incomplete elliptic integrals of the first and second kinds respectively.








\end{document}